# Integration of roadside camera images and weather data for monitoring winter road surface conditions

**Juan Carrillo**
**University of Waterloo**

**Mark Crowley**
**University of Waterloo**

## Abstract

During the winter season, real-time monitoring of road surface conditions is critical for the safety of drivers and road maintenance operations. Previous research has evaluated the potential of image classification methods for detecting road snow coverage by processing images from roadside cameras installed in RWIS (Road Weather Information System) stations. However, there are a limited number of RWIS stations across Ontario, Canada; therefore, the network has reduced spatial coverage. In this study, we suggest improving performance on this task through the integration of images and weather data collected from the RWIS stations with images from other MTO (Ministry of Transportation of Ontario) roadside cameras and weather data from Environment Canada stations. We use spatial statistics to quantify the benefits of integrating the three datasets across Southern Ontario, showing evidence of a six-fold increase in the number of available roadside cameras and therefore improving the spatial coverage in the most populous ecoregions in Ontario. Additionally, we evaluate three spatial interpolation methods for inferring weather variables in locations without weather measurement instruments and identify the one that offers the best tradeoff between accuracy and ease of implementation.

## Resume

Pendant la saison hivernale, la surveillance en temps réel de l'état de la chaussée est essentielle pour la sécurité de conduite et des opérations d'entretien des routes. Des recherches antérieures ont évalué le potentiel des méthodes de classification des images pour détecter la couverture neigeuse sur les routes en traitant des images provenant de caméras routières installées sur des stations RWIS (système d'information météorologique routière). Cependant, il s'agit d'une tâche difficile en raison de limitations telles que la résolution de l'image, l'angle de la caméra et l'éclairage. Dans cette étude, nous proposons l'intégration d'images et de données météorologiques recueillies à partir des stations RWIS avec des images provenant d'autres caméras routières du MTO (Ministère des Transports d'Ontario) et des données météorologiques provenant des stations d'Environnement Canada. De plus, nous utilisons des statistiques spatiales pour quantifier les avantages de l'intégration des trois jeux de données dans le sud de l'Ontario, et partant, l'amélioration de la couverture spatiale dans les écorégions les plus densément peuplées de l'Ontario. De plus, nous avons évalué trois méthodes d'interpolation spatiale climatique pour identifier celle qui offre le meilleur équilibre entre précision et facilité de mise en œuvre.

# 1. INTRODUCTION

Vision Zero is an innovative approach targeted to improve road safety to the point where zero fatalities happen in the roads [1]. It was created by the Swedish Government in 1997 and encompasses strategies where multiple stakeholders, such as Transportation Offices, Policy makers, car makers, and road users, just to name a few; work simultaneously toward reducing the factors related to road accidents. International and national organizations across the world have been promoting the Vision Zero approach over the past few years, with significant and quantified benefits in cities of developed and developing countries [2]. The main differences between conventional road safety plans and this new approach is a strong focus on engaging all the stakeholders in the planning process and a data-driven foundation that leverages innovative technologies toward embedding safety on every stage of the planning and operation of road infrastructure [3].

Several cities in Canada have already launched their Vision Zero plans and are working actively on promoting citizen engagement and encouraging the use of state-of-the-art technologies to materialize the goal of zero fatalities on the road. For instance, the Vision Zero plan in Toronto identifies areas with high priority to improve road safety, divided into five road user segments: motorcyclists, cyclists, pedestrians, older adults, and school children. They also identify two dangerous practices: aggressive driving and driver distraction [4]. In terms of the promotion of data-driven technologies, the City of Toronto ran a public Challenge in 2018 where citizens, universities, and the general public were invited to design data-driven innovations to help materialize the Vision Zero plan. While public engagement and education are both contributing factors toward improving road safety, engineering innovations in the planning and operation of road infrastructure are also crucial for implementing road safety into transportation infrastructure. In particular, the use of Big Data and Artificial Intelligence technologies within transportation systems is becoming more and more frequent. Specific applications such as video analytics are instrumental in enhancing traffic operations, monitoring road conditions, and producing valuable insights to better understand risk factors [5].

Real-time monitoring of road surface conditions using cameras is critical for the safety of drivers and road maintenance operations. A study conducted by the Iowa Department of Transportation, USA, confirmed that the use of roadside cameras for snow removal operations can reduce the number of road patrol expeditions by 33% and the overall cost of those expeditions by 14% [6]. Additionally, a study led by the Minnesota Department of Transportation, USA, found that dash and ceiling-mounted cameras on snowplows allow for streamlined operations and better communication of road conditions across Transportation offices, contractors, and the general public [7].

In that regard, advanced image classification methods for classifying road snow coverage by processing images from roadside cameras have been introduced in recent research [8] [9]. However, it is still a challenging task due to limitations such as image resolution, camera angle, and illumination. Two common approaches to improve the accuracy of image classification methods are: adding more input features to the model and increasing the number of samples in the training dataset.

In Ontario, the Ministry of Transportation (MTO) monitors winter road surface conditions through the Road Weather Information System (RWIS), a network of stations that include roadside cameras and specialized instruments to measure weather variables [10]. However, there are a limited number of RWIS stations across Ontario; therefore, the network has reduced spatial coverage. In order to enhance the range of the RWIS system, more roadside cameras and weather stations would be required. With that requirement in mind, the first objective of this study is to introduce a novel data integration approach that makes use of all the other MTO roadside cameras as well as Environment Canada weather stations to have both more images and more weather records for the purpose of road surface monitoring during the winter.

We highlight the potential of integrating additional datasets to improve the spatial coverage and accuracy for monitoring winter road surface conditions. Consequently, we examine the main characteristics of the input datasets, with a focus on the location and spatial configuration of the observing stations. Moreover, we also quantify the benefits of integrating the input datasets into a larger one that offers wider spatial coverage and more input features.

## 2. DATA INTEGRATION ANALYSIS

The area of interest for this study is the Province of Ontario in Canada. More specifically, we calculate the benefits of the data integration approach over the three most densely inhabited ecoregions of Ontario (Table 1.).

| Ecoregion | Population density inhabitants/km² | Rank across Canada |
|---|---|---|
| Lake Erie Lowland | 344 | 2nd |
| St. Lawrence Lowlands | 179 | 3rd |
| Manitoulin-Lake Simcoe | 66 | 6th |

**Table 1. The three most densely inhabited ecoregions in Southern Ontario, StatCan 2016.**

The RWIS system is the one currently used by the Ministry of Transportation of Ontario to monitor road winter surface conditions. Each RWIS station records images from roadside camera and weather variables using specialized measurement instruments. To improve the spatial coverage of the current system we propose to include all the other MTO roadside cameras as well. However, those MTO cameras do not record any weather variables. Therefore, we also propose the use of data from Environment Canada stations to interpolate weather variables for each one of the added MTO cameras. Figure 1 shows the spatial configuration of the observing stations in the three input datasets and locates the three most populous ecoregions in Ontario.

The reason why ecoregions become important to the analysis is twofold: on the one hand having more observations within the same ecoregion facilitates the interpolation of weather variables since we could expect similar weather conditions, on the other hand, we focus our analysis in the three most populous ecoregions in Canada; therefore maximizing the benefits of the system integration for a higher number of road users.

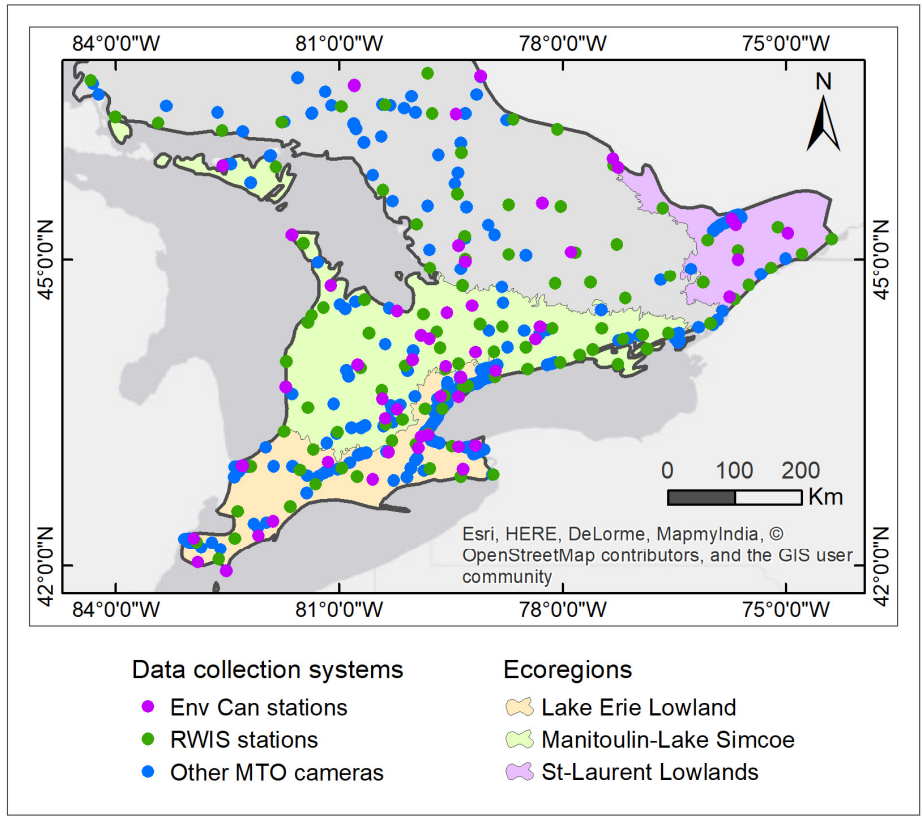

**Figure 1. Location of observing stations from the three input systems.**

Adding other MTO cameras to the RWIS system increases the total number of available roadside cameras in Ontario by more than four times and reduces the average distance to the nearest camera (Nearest neighbor NN) from around 38 km to less than 10 km. It also increases the number of cameras in the three most populous ecoregions in the province increases by more than six times. (Table 2.)

| Type | # of locations in Ontario | Avg. distance to NN (km) | # of locations in three populous ecoregions |
|---|---|---|---|
| RWIS | 139 | 38.4 | 68 |
| Other MTO | 439 | 7.2 | 364 |
| RWIS + MTO | 578 | 9.4 | 432 |

**Table 2. Adding other MTO roadside cameras to increase the number of images.**

The size (in pixels) and format of the images (.jpg) collected by RWIS and other MTO cameras is generally the same. In addition, every camera in both systems takes at least one picture of the road every 15 minutes, which is enough for monitoring winter road surface in most cases. A visual inspection of images taken by more than 40 RWIS stations and 30 other MTO cameras confirms that the point of view with respect to the road and the viewing angle of the roadside cameras in both systems are very similar; a fact that facilitates the combined processing of all the images.

Highway 15, near Otter Lake. RWIS ER-17.    QEW West of Fifty Rd.

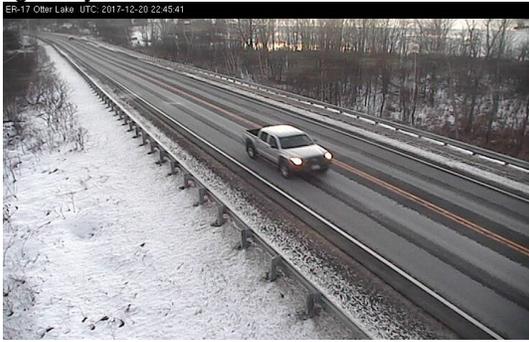 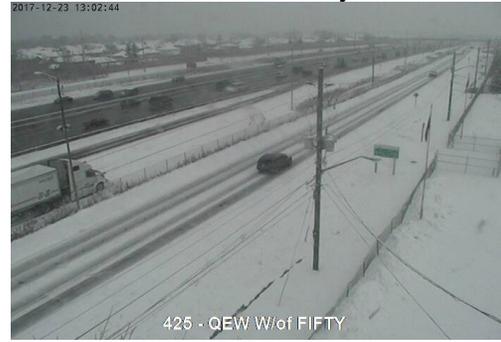

**Figure 2. On the left, an image from an RWIS station. On the right an image from an MTO camera.**

Moreover, adding weather stations from Environment Canada to the RWIS system increases the number of weather stations in Ontario by 1.7x and reduces the average distance to the nearest weather station (Nearest neighbor NN) from around 38 km to less than 26 km. More importantly, the number of stations in the three most populous ecoregions in the province increases by 1.7x. (Table 3.).

| Type | # of locations in Ontario | Avg. distance to NN (km) | # of locations in three populous ecoregions |
|---|---|---|---|
| RWIS | 139 | 38.4 | 68 |
| Env. Canada | 99 | 35.8 | 45 |
| RWIS + Env. Can | 238 | 25.7 | 113 |

**Table 3. Adding Environment Canada stations to interpolate weather data.**

Having more weather stations allows us to interpolate weather variables for all other MTO roadside cameras by using observations from both the RWIS system and the Environment Canada weather network. Furthermore, in order to better understand the spatial configuration of all three input datasets and to investigate the performance of some selected spatial interpolation methods, we conduct a more comprehensive analysis supported by spatial statistics.

In the following step of our analysis, we study the spatial clustering of observing locations in the three systems: RWIS, other MTO cameras, and Environment Canada stations. In other words, we evaluate how clustered the stations are in those three systems under the assumption that weather stations distributed at approximately random locations generally produce better weather interpolation compared to settings where stations are clustered [11].

For this purpose, we select the L-Function [12] to identify whether the spatial distribution of stations in each system corresponds to a completely random spatial point pattern or in contrary if the distribution corresponds to a clustered point pattern. The L-Function works by comparing the spatial distribution of the points in the reference sets (RWIS, MTO, EnvCan) against a set of points generated at random locations by sampling coordinates as uniformly distributed random variables within the area of study. What makes the L-Function advantageous for our analysis is the fact that it repeats the described process for multiple scales; therefore, it outputs a function that allows us to infer the degree of spatial randomness of every input dataset at different distance bands.

Figure 3 shows how clustered or randomly distributed are each one of the three datasets when seen across multiple spatial scales, in other words, when compared against multiple sets of randomly generated points each one having an increasing average distance between the random points. When the red line (reference set) is above the blue line (random sets of points), as we see for the RWIS dataset (a) and the other MTO cameras dataset (b) we infer that both datasets are highly clustered, which makes sense considering those cameras are all installed beside major roads. For Environment Canada stations (c), the red and blue lines are close to each other, which means those stations are distributed randomly across Ontario. In all three cases, we also plot a confidence interval obtained by creating nine permutations of the random points.

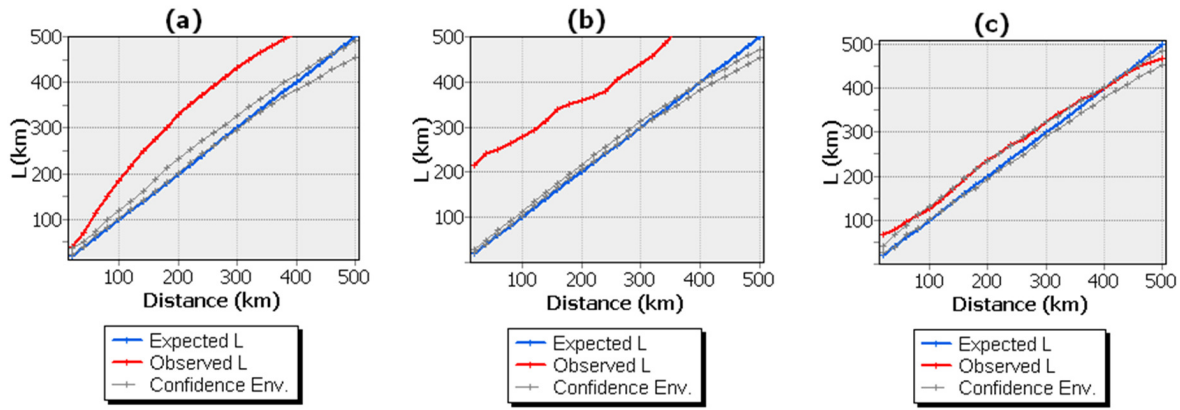

**Figure 3. Multi-distance spatial cluster (L-Function) plots for: (a) RWIS stations, (b) other MTO cameras, and (c) Environment Canada stations.**

Based on previous literature related to the optimal location of weather stations [11][13][14][15], we infer that the random spatial distribution of Environment Canada stations improves the interpolation of weather variables across Ontario compared to the clustered locations of the RWIS stations. Therefore, when combined with RWIS stations we can increase the spatial coverage and have a more distributed set of input stations to interpolate weather variables for MTO cameras.

## 3. RESULTS

We compare the performance of three spatial statistics methods to interpolate three key weather variables, namely air temperature, wind speed, and air pressure. Before comparing the interpolation methods, in Table 4 we can see the summary statistics for the three weather variables of interest on two different dates; there was no snow in the first date and snow in the second.

| Summary statistics | T1 - No snow - 2017/11/07 08:00 | | | T2 - Snow - 2017/12/25 08:00 | | |
|---|---|---|---|---|---|---|
| | air temp. (°C) | wind speed (km/h) | pressure (kPa) | air temp. (°C) | wind speed (km/h) | pressure (kPa) |
| **Mean** | -1.921 | 4.912 | 99.950 | -12.186 | 13.587 | 98.518 |
| **Std. dev.** | 5.195 | 6.419 | 2.809 | 9.509 | 11.128 | 2.782 |
| **CV%** | ---------- | 131% | 3% | ---------- | 82% | 3% |

**Table 4. Summary statistics of three weather variables for a no-snow day and a snowy day.**

We see that the mean air temperature decreases by approximately 10°C and the mean wind speed triples from the no-snow to the snowy day; in contrary, air pressure only shows a slight change of less than 2%. Therefore, we infer a better correlation between air temperature and wind with road snow coverage. A more comprehensive analysis regarding the correlation between those weather variables and road snow coverage is beyond the scope of our study.

The three methods we compare are Inverse Distance Weighted (IDW), Radial Basis Function (RBF), and Ordinary Kriging (OK). As input data, we selected a sample of 40 RWIS stations and 40 Environment Canada stations, for those 80 stations we obtain data values for a no-snow day in November 2017, and for a snowy day in December 2017. Table 5 summarizes the performance of the three interpolation methods considered. In general, the Root Mean Square (RMS) tends to be higher for the snowy day, likely due to high spatial variability not represented by the conventional standard deviation statistic in Table 4. Ordinary Kriging scores the lowest RMS in four of the six day-variable pairs; however, it also requires significantly more time to set up due to a greater number of parameters and model design decisions. Consequently, we suggest RBF as the interpolation method that offers the best tradeoff between complexity and accuracy.

| Interpolation Method | T1 - No snow - 2017/11/07 08:00 | | | T2 - Snow - 2017/12/25 08:00 | | |
| --- | --- | --- | --- | --- | --- | --- |
| | Air temp. (°C) | Wind speed (km/h) | Pressure (kPa) | Air temp. (°C) | Wind speed (km/h) | Pressure (kPa) |
| **IDW** | 2.054 | 6.073 | 3.094 | 4.139 | 8.761 | 3.053 |
| **RBF** | 1.971 | 6.156 | 3.001 | **3.898** | 8.718 | **2.963** |
| **Ord. Kriging** | **1.868** | **5.660** | **2.992** | 3.921 | **8.654** | 2.999 |

**Table 5. Root Mean Square of three interpolation methods applied on a no-snow and snowy day.**

In all three interpolation methods, we set up three and six, as the minimum and the maximum number of neighbors to use as input for the calculations. All parameter values are optimized through cross-validation, except for Ordinary Kriging where we set up the parameters based on findings from a study targeted to determine the optimal location of RWIS stations across multiple regions in North America [16].

Specific parameter configurations for each of the three interpolation methods are:
- IDW: Optimized power parameter
- RBF: Completely regularized spline with optimized kernel parameter.
- OK: First order trend removal with the exponential kernel. For the experimental variogram, we set up a lag size of 10 km with 20 lags. The selected semivariogram is always Gaussian with a fixed range of 100km.

By adding all other MTO cameras as image data sources, the total number of cameras in the combined dataset increases from 139 to 578 across Ontario and the average distance to the nearest camera decreases from 38.4km to 9.4km. Additionally, six times more cameras are available in the three most populated ecoregions in Ontario. Moreover, the experimental evaluation of three spatial interpolation methods for inferring weather variables in unobserved locations shows that the best tradeoff between complexity and accuracy is offered by Radial Basis Functions (RBF). Overall, we introduced a novel data integration approach to improve the spatial coverage of winter road surface monitoring stations and provide experimental evidence of the benefits, especially for the most densely populated areas in Southern Ontario.

## 4. DISCUSSION

Generally, RWIS stations also include pavement-embedded sensors to better determine the conditions of the road, especially the sub-surface temperature. Even though our analysis does not consider these sensors, the benefits of our suggested data integration approach are still significant considering that our goal is to facilitate the maintenance operations by providing a wider coverage for weather and camera data only. Further research can look at optimal ways to integrate data from pavement-embedded sensors as a complementary data source.

From the perspective of government transportation offices, our approach can provide actionable insights which can be used to more selectively perform manual patrolling to better identify road surface conditions. From a broader perspective, integrating these three datasets is feasible and can benefit the design and development of automated image classification methods for monitoring road snow coverage. Which in turn can help materialize the Vision Zero by improving the road maintenance operations and reducing the number of incidents due to poor road conditions during the winter.

## 5. CONCLUSIONS

To summarize, as part of this data integration study we first quantify the benefits of extending the RWIS system with images from other MTO cameras and weather observations from Environment Canada. Then we use spatial statistics to better understand the spatial configuration of stations in the three systems and find that Environment Canada stations provide a better spatial coverage than existing RWIS stations. Furthermore, we evaluate three different weather interpolation methods under snowy and no-snow weather conditions and suggest RBF as the one that better balances accuracy and ease of implementation.

Our initial results are promising and demonstrate that additional image and weather datasets can be incorporated to road monitoring systems, leading to measurable improvements in road monitoring tasks.